\documentclass[runningheads]{llncs}

\usepackage[utf8]{inputenc} 
\usepackage[T1]{fontenc}    
\usepackage{hyperref}       
\usepackage{url}            
\usepackage[anythingbreaks]{breakurl}

\usepackage{breakurl}
\usepackage{hyperref}
\usepackage{booktabs}       
\usepackage{amsfonts}       
\usepackage{nicefrac}       
\usepackage{microtype}      
\usepackage{lipsum}
\usepackage{graphicx}
\usepackage{amsmath}
\usepackage[table]{xcolor}
\usepackage[square,numbers]{natbib}
\usepackage{color}
\newcommand{\bs}{\boldsymbol}

\newcommand{\selector}{{\ensuremath \kappa}}

\newcommand{\bstheta}{{\ensuremath{\boldsymbol{\theta}}}}

\title{\bf Sweeter than SUITE: 
Supermartingale Stratified Union-Intersection Tests of Elections}

\titlerunning{Sweeter than SUITE}

\author{ 
 Jacob V. Spertus 
 \and
 Philip B. Stark\thanks{Authors listed alphabetically.}
}

\institute{
University of California, Berkeley, 
Department of Statistics\\
\email{jakespertus@berkeley.edu};
\email{pbstark@berkeley.edu}
}

\authorrunning{
J.V. Spertus and P.B. Stark
}

\begin{document}
\maketitle

\begin{abstract}
  Stratified sampling can be useful in risk-limiting audits (RLAs), for instance, to accommodate heterogeneous voting equipment or laws that mandate jurisdictions draw their audit samples independently.
  We combine the union-intersection tests in SUITE, the reduction of RLAs to testing whether the means of a collection of lists are all  $\leq 1/2$
  of SHANGRLA, and the  \textit{nonnegative supermartingale} (NNSM) tests in ALPHA
  to improve the efficiency and flexibility of stratified RLAs.
  A simple, non-adaptive strategy for combining stratumwise NNSMs decreases the measured risk in the 2018 pilot hybrid audit in Kalamazoo, Michigan, USA by more than an order of magnitude, from 0.037 for SUITE to 0.003 for our method.
  We give a simple, computationally inexpensive, adaptive rule for deciding which stratum to sample next that reduces audit workload by as much as 74\%
  in examples. 
  We also present NNSM-based tests that are computationally tractable even when there are many strata, illustrated with a simulated audit stratified across California's 58~counties.
    \keywords{risk-limiting audit, election integrity, supermartingale test, intersection hypothesis, multi-armed bandit}
\end{abstract}


\section{Introduction}

Most U.S.\ jurisdictions use computers to tabulate votes. 
Like all computers, vote tabulators are vulnerable to bugs, human error, 
and deliberate malfeasance---a fact that has been exploited (rhetorically, if not in reality) to undermine trust in U.S.\ elections \citep{levine20, chaitlin20, kahn20, baker20}.

To deserve public trust, elections must be trustworthy,
despite relying on untrustworthy software, hardware, and people: they should provide convincing affirmative evidence that the reported winners really won \citep{starkWagner12, appelEtal20, appelStark20}. 
Risk-limiting audits (RLAs) are a useful tool for conducting such \emph{evidence-based elections}.
RLAs have a specified maximum chance---the \textit{risk limit} $\alpha$---of not correcting the reported outcome if it is wrong, and never change the reported outcome if it is correct. 
Below we present methods to reduce the number of ballots that must be manually inspected in an RLA
when the reported outcomes are correct, for stratified audit samples.

In a ballot-level \textit{comparison} RLA, manual interpretations of the votes on randomly sampled ballot cards are compared to their corresponding \textit{cast vote records} (CVRs), the  system's interpretation of the votes on those cards. 
In a \textit{ballot-polling} RLA, votes are read manually from
randomly selected cards, but those votes are not compared to the system's
interpretation of the cards. 
All else equal, ballot-level comparison RLAs are more efficient than ballot-polling RLAs, but they require
the voting system to export CVRs in a way that the corresponding card can be uniquely identified.
Not all voting systems can.

Stratified random sampling 
can be mandatory or expedient
in RLAs. 
Some states' laws require audit samples to be drawn independently across jurisdictions
 (e.g., California Election Code \S~336.5 and \S~15360), in which case the audit sample for any contest that crosses jurisdictional boundaries is stratified. 
Stratifying on the technology used to tabulate votes can increase efficiency by allowing \textit{hybrid audits} \citep{ottoboniEtal18,howardEtal19}, which use ballot-level comparison
in strata where the voting technology supports it and ballot-polling elsewhere. 
Another reason to use stratification is to
allow RLAs to start before all ballots have been tabulated \citep{stark19b}.

The next section briefly reviews prior work on stratified audits.
Section~\ref{sec:mathematics} introduces notation and stratified risk measurement, then presents our improvements: (i)~sharper $P$-values from new risk-measuring functions; (ii)~sequential stratified sampling that adapts to the observed data in each stratum to increase efficiency; and (iii)~a computationally efficient method for an arbitrary number of strata.
Section~\ref{sec:simulations} evaluates the innovations using case studies and simulations. 
Section~\ref{sec:discussion} discusses the results and gives recommendations for practice.

\section{Past Work}
\label{sec:past_work}
The first RLAs involved stratified batch comparison, using the maximum error across strata and contests as the test statistic \citep{stark08a,stark08d,stark09a,hallEtal09}, a rigorous but inefficient approach.
\citet{higginsEtal11} computed sharper $P$-values for the same test statistic
using dynamic programming. 
SUITE \citep{ottoboniEtal18,howardEtal19} uses \textit{union-intersection tests} to represent the null hypothesis that one or more reported winners actually lost as a union of intersections of hypotheses about individual strata;
it involves optimization problems that are hard to solve when there are more than two strata. 

More recently, SHANGRLA \citep{stark20a} has reduced RLAs to a canonical form: testing whether the means of finite, bounded lists of numbers (representing ballot cards) are all less than 1/2, which allows advances in statistical inference about bounded populations to be applied directly to RLAs. 
\citet{stark20a} showed that union-intersection tests can be used 
with SHANGRLA
to allow \textit{any} risk-measuring function to be used in any stratum in stratified audits.

\citet{stark22} provided a new approach to union-intersection tests using nonnegative supermartingales (NNSMs): \textit{intersection supermartingales},
which open the possibility of reducing sample sizes by adaptive \textit{stratum selection} (using the first $t$ sampled cards to select the stratum from which to draw the $(t+1)$th card). 
\citet{stark22} does not provide an algorithm for stratum selection or evaluate the performance of the approach; this paper does both.

\section{Stratified audits}
\label{sec:mathematics}
We shall formalize stratified audits using the SHANGRLA framework \citep{stark20a}, which unifies comparison and polling audits. 
We then show how to construct a stratified comparison audit using SHANGRLA, how to measure the risk based on a stratified sample, and how adaptive sequential stratified sampling can improve efficiency. 

\subsection{Assorters and assertions}

Ballot cards are denoted $\{b_i\}_{i=1}^N$.
An assorter $A$
assigns a number $A(b_i) \equiv x_i \in [0,u]$
to ballot card $b_i$ \citep{stark20a} and the value $A(c_i)$ to CVR $i$.
The value an assorter assigns to a card
depends on the votes on the card,
the social choice function, and possibly on the machine interpretation of that card and others (for comparison audits).
\citet{stark20a} describes how to define a set of assorters for many social choice functions (including majority, multiwinner majority, supermajority, Borda count, approval voting, all scoring rules, D'Hondt, STAR-Voting, and IRV) such that the reported winner(s) really won if the mean of every assorter in the set is greater than $1/2$.
The claim that an assorter mean is $> 1/2$ is called an \textit{assertion}.
An RLA with risk limit $\alpha$ confirms the outcome of a contest if it rejects the \textit{complementary null} that the assorter mean is $\le 1/2$ at significance level $\alpha$ for every assorter relevant to that contest. 

In a stratified audit, the population of ballot cards is partitioned into $K$ disjoint \emph{strata}.
Stratum $k$ contains $N_k$ ballot cards, so $N = \sum_k N_k$.
The \textit{weight} of stratum $k$ is $w_k := N_k/N$;
the weight vector is 
$\bs{w} := [w_1,...,w_K]^T$.
For each assorter $A$ there is a set of
assorter values $\{x_i\}_{i=1}^N$.
Each assorter may have its own upper bound $u_k$ in stratum $k$.\footnote{The notation we use does not allow $u$ to vary by draw, but the theory in \citet{stark22} permits it, and it is useful for batch-comparison audits.} 
The true mean of the assorter values in stratum $k$ is $\mu_k$; 
$\bs{\mu} := [\mu_1,...,\mu_K]^T$.
The overall assorter mean is
$$\mu := \frac{1}{N} \sum_{i=1}^N x_i = \sum_{k=1}^K \frac{N_k}{N} \mu_k = \bs{w}^T \bs{\mu}.$$
Let $\bs{\theta} = [\theta_{1},...,\theta_K]^T$ with $0 \le \theta_k \leq u_k$.
A single \emph{intersection null} is of the form
$\bs{\mu} \le \bs{\theta}$, i.e., $\cap_{k=1}^K \{\mu_k \le \theta_k\}$. 
The 
\textit{union-intersection form} of the \emph{complementary null} that the outcome is incorrect is:
\begin{equation}
H_0: \bigcup_{\bs{\theta}: \bs{w}^T \bs{\theta} \leq \frac{1}{2}} \bigcap_{k=1}^K \{\mu_k \leq \theta_k \}. \label{eqn:union-intersection_null}
\end{equation}

From stratum $k$ we have $n_k$ samples $X^{n_k}_k := \{X_{1k},...,X_{n_k k}\}$ drawn by simple random sampling, with or without replacement, independently across strata. 
Section~\ref{sec:union_intersection_tests}
shows how to use single-stratum hypothesis tests (of the the null $\mu_k \le \theta_k$) to test (\ref{eqn:union-intersection_null}). 
First, we show how to write stratified comparison audits in this form.

\subsection{Stratified comparison audits}
\label{sec:stratified_comparison}
In SHANGRLA, comparison audits involve translating the original assertions about the true votes into  assertions about the reported results and discrepancies between the true votes and the machine's record of the votes \citep[Section~3.2]{stark20a}.
For each assertion, the
corresponding 
\emph{overstatement assorter} assigns ballot card $b_i$ a bounded, nonnegative number that depends on the votes on that card, that card's CVR, and the reported results.
The original assertion is true if the average of the overstatement assorter values is greater than 1/2.

We now show that for stratified audits, the math is simpler if, as before, we assign a nonnegative number to each card
that depends on the votes and reported votes, but instead of comparing the average of the resulting list to 1/2, we compare it to a threshold that depends on the hypothesized stratum mean $\theta_k$.

Let $u^A_k$ be the upper bound on the original assorter for stratum $k$ and $\omega_{ik} := A(c_{ik}) - A(b_{ik}) \in [-u^A_k, u^A_k]$ be the \emph{overstatement} for the $i$th card in stratum $k$, where $A(c_{ik})$ is the
value of the assorter applied to the CVR and $A(b_{ik})$ is the  value of the assorter for the true votes on that card.
Let $\bar{A}^b_k$, $\bar{A}^c_k$, and $\bar{w}_k = \bar{A}^c_k - \bar{A}^b_k$ be the true assorter mean, reported assorter mean, and average overstatement, all for stratum $k$. 

For a particular $\bs{\theta}$, the intersection null claims that in stratum $k$, $\bar{A}^b_k \leq \theta_k$.
Adding $u_k^A-\bar{A}_k^c$ to both sides of the inequality yields 
$$u^A_k - \bar{\omega}_k \leq \theta_k + u^A_k - \bar{A}^c_k.$$ 
Letting $u_k := 2 u^A_k$, take $B_{ik} := u^A_k - \omega_{ik} \in [0, u_k]$ and $\bar{B}_k := \frac{1}{N_k} \sum_{i=1}^{N_k} B_{ik}$. 
Then $\{B_{ik}\}$ is a bounded list of nonnegative numbers, and the assertion in stratum $k$ is true if $\bar{B}_k > \beta_k := \theta_k + u^A_k - \bar{A}_k^c$,
where all terms on the right are known.
Testing whether $\bar{B} \le \beta_k$ is the canonical problem solved by ALPHA \citep{stark22}.
The intersection null
can be written
$$\bar{B}_k \leq \beta_k ~~\mbox{for all}~~ k \in \{1,\ldots,K\}.$$
Define $\bs{u} := [u_1, \ldots, u_K]^T$. As before, we can reject the complementary null if we can reject \textit{all} intersection nulls 
$\bs{\theta}$ for which $\bs{0} \le \bs{\theta} \le \bs{u}$ and $\bs{w}^T \bs{\theta} \leq 1/2$.

\subsection{Union-intersection tests}
\label{sec:union_intersection_tests}

A union-intersection test for (\ref{eqn:union-intersection_null}) combines evidence across strata
to see whether any intersection null in the union is plausible given the data, that is, to check whether the $P$-value of any intersection null in the union is greater than the risk limit.

Consider a fixed vector $\bs{\theta}$ of within-stratum nulls.
Let $P(\bs{\theta})$ be a valid $P$-value for the intersection null $\bs{\mu} \leq \bs{\theta}$.
Many functions can be used to construct $P(\bs{\theta})$ from
tests in individual strata; two are presented below.
We can reject the union-intersection null (\ref{eqn:union-intersection_null}) if we can reject the intersection null for all feasible $\bs{\theta}$ in the half-space $\bs{w}^T \bs{\theta} \leq 1/2$.
Equivalently, $P(\bs{\theta})$ maximized over feasible $\bs{\theta}$ is a $P$-value for (\ref{eqn:union-intersection_null}):
\begin{align*}
    P^* := \max_{\bs{\theta}}~ \{ P(\bs{\theta}):
    \boldsymbol{0} \le 
    \bs{\theta} \le \bs{u} ~\mbox{ and }~
    \boldsymbol{w}^T \bs{\theta} \leq {1}/{2}\}.
\end{align*}
This method is fully general in that it can construct a valid $P$-value for (\ref{eqn:union-intersection_null}) from stratified samples and any mix of risk-measuring functions that are individually valid under simple random sampling. 
However, the tractability of the optimization problem depends on the within-stratum risk-measuring functions and the form of $P$ used to pool risk. 
So does the efficiency of the audit.

We next give two valid combining rules $P(\bs{\theta})$.
Section~\ref{sec:p_vals}presents some choices for within-stratum risk measurement to construct
$P(\bs{\theta})$.

\subsection{Combining Functions}

\citet{ottoboniEtal18} and \citet{stark20a} 
calculate $P$ for the intersection null
using Fisher's combining function. 
Let $p_k(\theta_k)$ be a $P$-value for the single-stratum null $H_{0k}: \mu_k \leq \theta_k$.
Define the pooling function
$$P_F(\bs{\theta}) := 1 -  \chi^2_{2K} \left ( -2 \sum_{k=1}^K \log p_k(\theta_k) \right),$$
where $\chi^2_{2K}$ is the CDF of the chi-squared distribution with 2K degrees of freedom. 
The term inside the CDF, $-2 \sum_{k=1}^K \log p_k(\theta_k)$, is Fisher's combining function\footnote{%
   Other combining functions could be used, including Liptak’s or Tippett's. See Chapter 4 of \citet{pesarinSalmaso10}}.
Because samples are independent across strata, $\{p_k(\theta_k)\}_{k=1}^K$ are
independent random variables, so Fisher's combining function is dominated by the chi-squared distribution with $2K$ degrees of freedom \citep{ottoboniEtal18}. 
The maximum over $\bs{\theta}$, $P_F^*$, is a valid $P$-value for (\ref{eqn:union-intersection_null}).

\subsection{Intersection supermartingales}

\citet{stark22} derives a simple form for the  $P$-value for an intersection null when supermartingales are used as test statistics within strata. 
Let $M_{n_k}^k(\theta_k)$ 
be a supermartingale constructed from $n_k$ samples drawn from stratum $k$ when the null $\mu_k \leq \theta_k$ is true. 
Then the product of these supermartingales is also a supermartingale under the intersection null, so its reciprocal (truncated above at 1) is a valid $P$-value \citep{stark22, waudby-smithEtal21}:
$$P_M(\bs{\theta}) := 1 \wedge \prod_{k=1}^K M_{n_k}^k(\theta_k)^{-1}.$$
Maximizing $P_M(\bs{\theta})$ (equivalently, minimizing the intersection supermartingale) yields $P_M^*$, a valid $P$-value for (\ref{eqn:union-intersection_null}).

 \subsection{Within-stratum $P$-values}
\label{sec:p_vals}
The class of within-stratum $P$-values that can be used to construct $P_F$ is very large, but $P_M$ is limited to functions that are supermartingales under the null. Possibilities include:
\begin{itemize}
    \item \textbf{SUITE}, which computes $P_F^*$ for two-stratum hybrid audits. 
    The $P$-value in the CVR stratum uses the MACRO test statistic \citep{stark09c}; the $P$-value in the no-CVR stratum takes a maximum over many values of Wald's SPRT indexed by a nuisance parameter representing the number of non-votes in the stratum. The maximations in MACRO and over a nuisance parameter in the SPRT make SUITE less efficient than newer methods based on SHANGRLA \citep{stark20a}.
    \item \textbf{ALPHA}, which constructs a betting supermartingale as in \citet{WaudbysmithRamdas20}, but with an alternate parameterization \citep{stark22}. Such methods are among the most efficient for RLAs \citep{waudby-smithEtal21, stark22}, but the efficiency depends on how the tuning parameter $\tau_{ik}$ is chosen. \citet{stark22} offers a sensible strategy based on setting $\tau_{ik}$ to a stabilized estimate of the true mean $\mu_k$. 
    We implement that  approach and a modification that is more efficient for comparison audits. Both $P_M^*$ and $P_F^*$ can be computed from stratum-wise ALPHA supermartingales. 
    However, finding the maximum $P$-value over the union is prohibitively slow when $K>2$. 
    \item \textbf{Empirical Bernstein} (EB), which is a supermartingale presented in \citet{Howard21} and \citet{WaudbysmithRamdas20}. Although they are generally not as efficient as ALPHA and other betting supermartingales \citep{WaudbysmithRamdas20}, EB supermartingales have an exponential analytical form that makes $\log P_M(\bs{\theta})$ or $\log P_F(\bs{\theta})$ linear or piecewise linear in $\bs{\theta}$. Hence, $P_M^*$ and $P_F^*$ can be computed quickly for large $K$ by solving a linear program.
\end{itemize}
We compare the efficiency of these risk-measuring functions in Sections~\ref{sec:stratum_selection} and \ref{sec:suite_comparison}.

\subsection{Sequential stratum selection}

The use of sequential sampling in combination with stratification presents a new possibility for reducing workload: sample more from strata that are providing evidence against the intersection null and less from strata that are not helping. 
To set the stage, suppose we are conducting a ballot-polling audit with two strata of equal size and testing the intersection null $\bs{\theta} = [0.25, 0.75]^T$. 
We have drawn 50 ballot cards from each stratum and found sample assorter means of $[0.5, 0.6]^T$. 
Given the data, it seems plausible that drawing more samples from the first stratum will strengthen the evidence that $\mu_1 > 0.25$, but additional sampling from the second stratum might not provide evidence that $\mu_2 > 0.75$: to reject the intersection null, it might help to draw disproportionately from the first stratum. 
Perhaps suprisingly, such adaptive sampling yields valid inferences when the $P$-value is constructed from supermartingales
and the stratum selection function depends only on past data. 
We now sketch why this is true.  

For $t \in \mathbb{N}$ and a particular vector of hypothesized 
stratum means $\bstheta$, let 
$$\selector_t(\bstheta) \in \{1,...,K\}$$
denote the stratum from which the $t$-th sample was drawn for testing the hypothesis $\bs{\mu} \leq \bstheta$.
We call $\selector(\bstheta) := (\selector_t(\bstheta))_{t \in \mathbb{N}}$ the \emph{stratum selector} for null $\bstheta$.
Crucially, $\selector(\bstheta)$ is a \textit{predictable sequence}
with respect to $(X_t)_{t \in \mathbb{N}}$
in the sense that $\selector_t(\bstheta)$ can 
depend on
$X^{t-1} := \{X_1,\ldots,X_{t-1}\}$ but not on $X_i$ for $i \ge t$; it 
could be deterministic given $X^{t-1}$ or may also depend on auxiliary randomness.   

For example, a stratum selector could ignore past data and select strata in a deterministic round-robin sequence or at random with probability proportional to stratum size.
Alternatively, a rule might select strata adaptively, for instance picking a stratum at random with probability proportional to the current value of each within-stratum supermartingale, so that strata with larger $M_{t_k}^k(\theta_k)$ are more likely to be chosen---an ``exploration--exploitation'' strategy.
In what follows we suppress the dependence on $\bstheta$ except when it is explicitly required for clarity.

Now, let 
$M_t^{\selector}(\bstheta) := \prod_{i=0}^t Z_i$ be
the test statistic for testing the null 
hypothesis that the vector of stratumwise
means is less than or equal to $\bstheta$. 
This is a supermartingale if the individual terms $Z_i$
satisfy a simple condition. 
Let $Z_0=1$ and $Z_i \geq 0$ for all $i$.
If 
\begin{equation} \label{eq:supermart-condition}
\mathbb{E}_\bstheta [Z_t | X^{t-1}] \le 1,
\end{equation}
then $(M_t^{\selector}(\bstheta))_{t \in \mathbb{N}_0}$ is a nonnegative
supermartingale starting at 1 under the null. 
By Ville's inequality \citep{ville39}, the thresholded inverse $(1 \wedge M_t^{\selector}(\bstheta)^{-1})_{t \in \mathbb{N}_0}$ is an anytime $P$-value sequence when $\bs{\mu} \leq \bs{\theta}$.

Condition (\ref{eq:supermart-condition}) holds if the $Z_i$ are terms extracted from a set of within-stratum supermartingales using a predictable stratum selector:
Let 
\begin{equation}
    \nu_t^\selector :=
    \# \{i \le t : \selector_i = \selector_t \}
\end{equation}
be the number of draws from stratum $k$ as of time $t$.
Suppose that for $k \in \{1, \ldots, K\}$,
$M_t^{k}( \theta_k) := \prod_{i=1}^t
Y_i^{k}(\theta_k)$
is a nonnegative supermartingale
starting at 1 when $X_{ik}$ is the
$i$th draw from stratum $k$ and the $k$th stratum mean is $\mu_k \leq \theta_k$.
Then if
\begin{equation}
    Z_i := Y_{\nu_i^\kappa}^{\selector_i}(\theta_{\selector_i}),
\end{equation}
condition~\eqref{eq:supermart-condition}
holds and the interleaved test statistic $M_t^{\selector}(\bstheta)$ is an intersection supermartingale under the null. 
We compare two stratum selection rules in Section~\ref{sec:stratum_selection}.

\section{Evaluations}
\label{sec:simulations}

\subsection{Combination and allocation rules}
\label{sec:stratum_selection}

We simulated a variety of two-stratum ballot-level comparison audits at risk limit $\alpha = 5\%$, with assorters defined as in Section~\ref{sec:stratified_comparison}.
The strata each contained $N_k = 1000$ ballot cards, all with valid votes.
Cards were sampled without replacement. 
The stratum-wise true margins were $[0\%, 20\%]$, $[0\%,10\%]$ or $[0\%,2\%]$,
corresponding to global margins of 10\%, 5\%, and 1\%, respectively.
Stratum-wise reported margins were also $[0\%, 20\%]$, $[0\%,10\%]$ or $[0\%,2\%]$, so error was always confined to the second stratum. 
Each reported margin was audited against each true margin in
300 simulations. 
Risk was measured by ALPHA or EB combined either as intersection supermartingales ($P_M^*$) or with Fisher's combining function ($P_F^*$), with one of two stratum selectors: proportional allocation or lower-sided testing.

In proportional allocation, the number of samples from each stratum is in proportion to the number of cards in the stratum. 
Allocation by lower-sided testing involves testing the null $\mu_k \geq \theta_{k}$ 
sequentially at level 5\% using the same 
supermartingale (ALPHA or EB) used to test the main (upper-sided) hypothesis of interest. 
This allocation rule ignores samples from a given stratum once the lower-sided hypothesis test rejects, since there is strong evidence that the null is true in that stratum. 
This ``hard stop'' algorithm is unlikely to be optimal, but it leads to a computationally efficient implementation and illustrates
the potential improvement in workload from adaptive stratum selection.

Tuning parameters were chosen as follows. 
ALPHA supermartingales were specified either with $\tau_{ik}$ as described in \citet[Section~2.5.2]{stark22} (ALPHA-ST, ``shrink-truncate'') or with a strategy that biases $\tau_{ik}$ towards $u_k$: (ALPHA-UB, ``upward bias''). 
The ALPHA-UB strategy 
helps in comparison audits because the distribution of assorter values consists of a  point mass at $u_A^k = u_k/2$ and typically small masses (with weight equal to the overstatement rates) at 0 and another small value.
This concentration of mass makes it advantageous to bet more aggressively that the next draw will be above the null mean;
that amounts to biasing $\tau_{ik}$ towards the upper bound $u_{k}$. 
Before running EB, the population and null were transformed to [0,1] by dividing by $u_k$. The EB supermartingale parameters $\lambda_{ik}$ were then specified following the ``predictable mixture'' strategy  \citep[Section~3.2]{WaudbysmithRamdas20}, truncated to be below $0.75$. Appendix~\ref{sec:appendix_computation} gives
more details of the ALPHA-ST and ALPHA-UB strategies and
the computations.

\begin{figure}[ht]
    \centering
    \includegraphics[width = \textwidth]{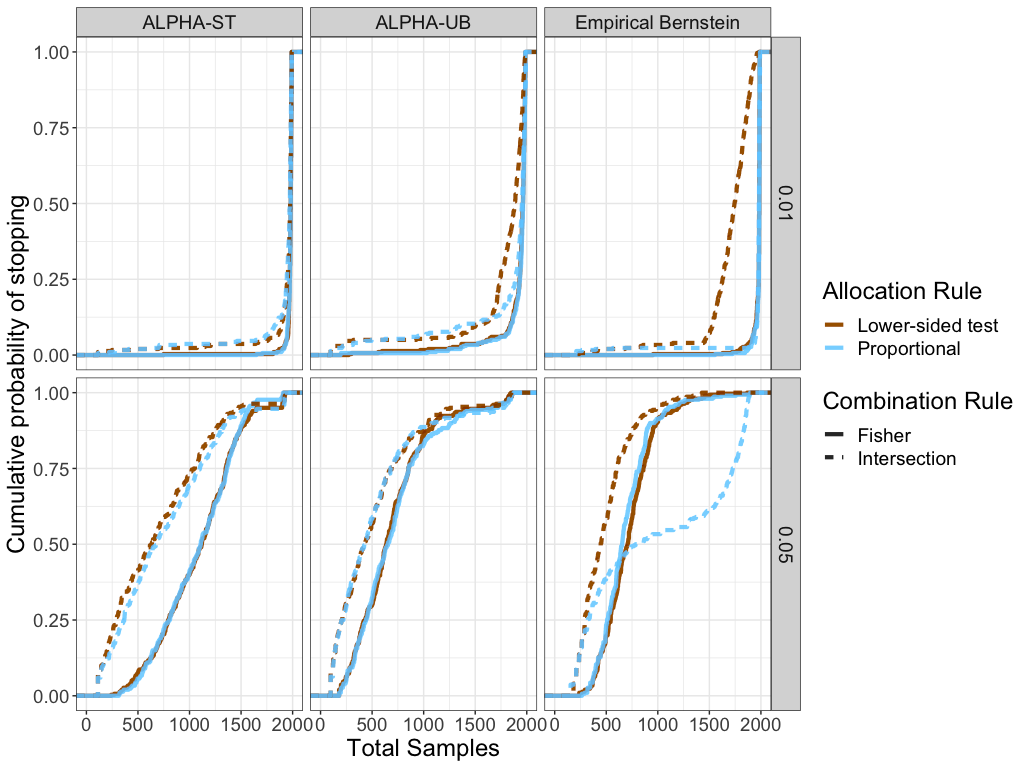}
    \caption{Probability that the audit will stop ($y$-axis) at or before different given sample sizes ($x$-axis) under different allocation rules (indicated by line color: orange for lower-sided testing and blue for proportional allocation)
    for different combining functions (indicated by line type: solid for Fisher's combining function and dashed for the intersection supermartingale) at risk limit $\alpha = 5\%$. 
    The true margins are in the rows (1\% or 5\%) while the reported margin is always 10\%. 
    Overstatement errors are confined to one stratum. ALPHA-ST = ALPHA with shrink-truncate $\tau_{ik}$; ALPHA-UB = ALPHA with $\tau_{ik}$ biased towards $u_{k}$.}
    \label{fig:allocation_power}
\end{figure}

Sample size distributions for some combinations of reported and true margins are plotted in Figure~\ref{fig:allocation_power} as 
(simulated) probabilities of stopping at or before a given sample size. Table~\ref{tab:stopping_times} gives estimated expected and 90th percentile sample sizes for each scenario and method.
Table~\ref{tab:workload_ratios} lists aggregate scores, computed by finding the ratio of the workload for each method over the smallest workload in each scenario, then averaging over scenarios by taking the geometric mean of these ratios.

Intersection supermartingales tend to dominate Fisher pooling
unless the stratum selector is chosen poorly (e.g., the bottom-right panel 
of Figure~\ref{fig:allocation_power} and the last row of Table~\ref{tab:workload_ratios}). 
Stratum selection with the lower-sided testing procedure is about as efficient as proportional allocation for the ALPHA supermartingales, but far more efficient than proportional allocation for EB.
The biggest impact of the allocation rule occurred for EB combined by intersection supermartingales when the reported margin was 0.01 and the true margin was 0.1: proportional allocation produced an expected workload of 752 cards, while lower-sided testing produced an expected workload of 271 cards---a 74\% reduction.
Table~\ref{tab:workload_ratios} shows that ALPHA-UB with intersection supermartingale combining and lower-sided testing is the best method overall;
ALPHA-UB with intersection combining and proportional allocation is a close second;
EB with intersection combining and lower-sided testing is also relatively sharp;
ALPHA-ST with Fisher combining is least efficient. 

We also ran simulations at risk limits 1\% and 10\%, which did not change the relative performance of the methods. 
However, compared to a 5\% risk limit, a 10\% risk limit requires counting about 17\% fewer cards and a 1\% risk limit requires about 38\% more, on average across scenarios and methods. 

\begin{table}[ht]
\centering
\scalebox{0.9}{
\begin{tabular}{c|c|c|crr|rr|rr}
  \hline
\textbf{Reported} & \textbf{supermartingale} & \textbf{Combination} & \textbf{Allocation} & \multicolumn{6}{c}{\textbf{True margin}} \\
\cline{5-10}
\textbf{margin} & & & \textbf{rule} & \multicolumn{2}{c}{\textbf{0.01}} & \multicolumn{2}{c}{\textbf{0.05}} & 
\multicolumn{2}{c}{\textbf{0.1}} \\
& & & & \textbf{Mean} & \textbf{90th} & \textbf{Mean} & \textbf{90th} & \textbf{Mean} & \textbf{90th} \\ 
  \hline
0.01 & ALPHA-ST & Fisher & Lower-sided test & 1970 & 1970 & 1011 & 1274 & 338 & 506 \\ 
   & & & Proportional & 1970 & 1970 & 1009 & 1274 & 338 & 540 \\
   \cline{3-10}
   & & Intersection & Lower-sided test & 1940 & 1940 & 558 & 848 & 181 & 284 \\ 
  &  &  & Proportional & 1940 & 1940 & 554 & 835 & 182 & 298  \\ 
  \cline{2-10}
  & ALPHA-UB & Fisher & Lower-sided test & 1402 & 1402 & 544 & 754 & 252 & 360  \\ 
  & &  & Proportional & 1402 & 1402 & 548 & 748 & 248 & 354 \\ 
   \cline{3-10}
  & & Intersection & Lower-sided test & 1106 & 1106 & 344 & \cellcolor{lime} 504 & 149 & 238 \\ 
  & &  & Proportional & 1106 & 1106 & \cellcolor{lime} 342 & 510 & \cellcolor{lime} 148 & \cellcolor{lime} 232 \\
  \cline{2-10}
  & Empirical Bernstein & Fisher & Lower-sided test & 1438 & 1438 & 649 & 768 & 384 & 498 \\ 
  &  &  & Proportional & 1438 & 1438 & 647 & 782 & 376 & 464 \\ 
   \cline{3-10}
  &  & Intersection & Lower-sided test & \cellcolor{lime} 1102 & \cellcolor{lime} 1102 & 478 & 652 & 271 & 378  \\ 
  & &  & Proportional & \cellcolor{lime} 1102 & \cellcolor{lime} 1102 & 982 & 1856 & 752 & 1728\\ 
  \hline
  0.05 & ALPHA-ST & Fisher & Lower-sided test & 1973 & 1986 & 908 & 908 & 305 & 426 \\ 
   & &  & Proportional & 1972 & 1984 & 908 & 908 & 298 & 412\\
    \cline{3-10}
   &  & Intersection & Lower-sided test & 1930 & 1980 & 428 & 428 & 145 & 212 \\ 
   & &  & Proportional & 1933 & 1982 & 428 & 428 & 151 & 228  \\ 
  \cline{2-10}
  & ALPHA-UB & Fisher & Lower-sided test & 1769 & 1970 & 428 & 428 & 217 & 292   \\ 
  & &  & Proportional & 1769 & 1972 & 428 & 428 & 217 & 288 \\ 
   \cline{3-10}
  & & Intersection & Lower-sided test & 1611 & 1884 & \cellcolor{lime} 256 & \cellcolor{lime} 256 & \cellcolor{lime} 122 & \cellcolor{lime} 176  \\ 
  & &  & Proportional & 1651 & 1962 & \cellcolor{lime} 256 & \cellcolor{lime} 256 & \cellcolor{lime} 122 & 180  \\ 
  \cline{2-10}
  & Empirical Bernstein & Fisher & Lower-sided test & 1882 & 1986 & 448 & 448 & 306 & 356 \\ 
  &  &  & Proportional & 1870 & 1986 & 448 & 448 & 304 & 354 \\ 
   \cline{3-10}
  &  & Intersection & Lower-sided test & \cellcolor{lime} 1610 & \cellcolor{lime} 1858 & 296 & 296 & 199 & 234 \\ 
  &  &  & Proportional & 1924 & 1982 & 296 & 296 & 302 & 376  \\
  \hline
  0.10 & ALPHA-ST & Fisher & Lower-sided test & 1971 & 1990 & 1088 & 1536 & 240 & 240 \\ 
  &  &  & Proportional & 1974 & 1990 & 1080 & 1509 & 240 & 240  \\
   \cline{3-10}
  &  & Intersection & Lower-sided test & 1910 & 1991 & 694 & 1312 & 112 & 112 \\ 
  &  &  & Proportional & 1894 & 1988 & 755 & 1347 & 112 & 112 \\ 
  \cline{2-10}
  & ALPHA-UB & Fisher & Lower-sided test & 1904 & 1984 & 696 & 1107 & 180 & 180 \\ 
  &  & & Proportional & 1914 & 1984 & 715 & 1263 & 180 & 180  \\ 
   \cline{3-10}
  &  & Intersection & Lower-sided test & 1756 & 1968 & 521 & 1046 & \cellcolor{lime} 98 & \cellcolor{lime} 98  \\ 
  & &  & Proportional & 1804 & 1990 & 534 & 1079 & \cellcolor{lime} 98 & \cellcolor{lime} 98 \\ 
  \cline{2-10}
  & Empirical Bernstein & Fisher & Lower-sided test & 1968 & 1988 & 716 & 987 & 238 & 238  \\ 
  &  &  & Proportional & 1974 & 1988 & 686 & 928 & 238 & 238 \\ 
   \cline{3-10}
  &  & Intersection & Lower-sided test & \cellcolor{lime} 1697 & \cellcolor{lime} 1901 & \cellcolor{lime} 487 & \cellcolor{lime} 799 & 154 & 154\\ 
  &  & & Proportional & 1939 & 1990 & 1000 & 1846 & 154 & 154 \\ 
   \hline
\end{tabular}
}
\caption{Expected and 90th percentile sample sizes for various risk-measurement functions, reported margins, and true margins, estimated
from 300 simulated audits at risk-limit $\alpha = 5\%$. 
The best result for each combination of reported margin, true margin, and summary statistic is highlighted.
Comparison audit sample sizes are deterministic when there is no error, so the expected value and 90th percentile are equal when the reported and true margins are equal.}
\label{tab:stopping_times}
\end{table}

\begin{table}[ht]
\centering
\begin{tabular}{l|l|lr}
  \hline
\textbf{supermartingale} & \textbf{Combination} & \textbf{Allocation} & \textbf{Score} \\ 
  \hline
ALPHA-ST & Fisher & Lower-sided test & 2.11 \\ 
   &  & Proportional & 2.10 \\ 
   \cline{2-4}
   & Intersection & Lower-sided test & 1.35 \\ 
   &  & Proportional & 1.37 \\ 
   \hline
  ALPHA-UB & Fisher & Lower-sided test & 1.47 \\ 
   &  & Proportional & 1.48 \\ 
   \cline{2-4}
   & Intersection & Lower-sided test & \cellcolor{lime} 1.01 \\ 
   &  & Proportional & 1.02 \\ 
   \hline
  Empirical Bernstein & Fisher & Lower-sided test & 1.73 \\ 
   &  & Proportional & 1.71\\ 
   \cline{2-4}
   & Intersection & Lower-sided test & 1.25 \\ 
   &  & Proportional & 1.78 \\ 
   \hline
\end{tabular}
\caption{Score for each method: the geometric mean of the expected workload over the minimum expected workload in each scenario. A lower score is better: a 1.00 would mean that the method always had the minimum expected workload. The best score is highlighted. A score of 2 means that workloads were twice as large as the best method, on average, across simulations and scenarios.}
\label{tab:workload_ratios}
\end{table}


\subsection{Comparison to SUITE}
\label{sec:suite_comparison}
SUITE was used in a pilot RLA of the 2018 
gubernatorial election in Michigan \citep{howardEtal19}. 
Three jurisdictions---Kalamazoo, Rochester Hills, and Lansing---were audited, but only Kalamazoo successfully ran a hybrid audit. 
We recalculated the risk on audit data from the closest race in Kalamazoo (Whitmer vs Schuette) 
using ALPHA with the optimized intersection supermartingale $P$-value $P_M^*$, ALPHA with the optimized Fisher $P$-value $P_F^*$, EB with $P_F^*$, 
and EB with $P_M^*$, and
compared these with the SUITE $P$-value.
Because we could not access the original order of sampled ballots in the ballot-polling stratum, we simulated $P$-values for 10,000 random ballot orders with the marginal totals in the sample.
We computed the mean, standard deviation, and 90th percentile of these $P$-values for each method.

To get the ALPHA $P$-values, we used ALPHA-UB in the CVR stratum and ALPHA-ST in the no-CVR stratum.
For EB $P$-values, we used the predictable mixture parameters of \citet{WaudbysmithRamdas20} to choose $\lambda_{ik}$, truncating at 0.75 in both strata.
Sample allocation was dictated by the original pilot audit:
8~cards from the CVR stratum (5,294 votes cast; diluted margin 0.55) and 32 from the no CVR stratum (22,732 votes cast; diluted margin 0.57).  

Table~\ref{tab:kalamazoo} presents $P$-values for each method. 
For ALPHA, the mean $P_F^*$ is about half the SUITE $P$-value;
for $P_M^*$, the mean is more than an order of magnitude smaller than the SUITE $P$-value. 
The $P$-value distributions for ALPHA
are concentrated near the mean.
On the other hand, 
the EB $P_M^*$ and $P_F^*$ $P$-values are both an order of magnitude larger than the SUITE $P$-value and their distributions are substantially more dispersed than the 
distributions of ALPHA $P$-values. 

\begin{table}[h]
    \centering
    \begin{tabular}{l|lll}
    \hline
         & \multicolumn{3}{c}{$\bs{P}$\textbf{-value}} \\
         \cline{2-4}
        \textbf{Method} & \textbf{Mean} & \textbf{SD} & \textbf{90th}\\
        \hline 
         SUITE &  0.037 & * & * \\
         ALPHA $P_F^*$ & 0.018 & 0.002 & 0.019  \\
         ALPHA $P_M^*$ & 0.003 & 0.000 & 0.003 \\
         EB $P_F^*$ & 0.348 & 0.042 & 0.390 \\
         EB $P_M^*$ & 0.420 & 0.134 & 0.561\\
         \hline
    \end{tabular}
    \caption{Measured risks ($P$-values) computed from the 2018 Kalamazoo MI audit data. 
    For SUITE, the original $P$-value is shown. For replications, the mean, standard deviation (SD), and 90th percentile of $P$-values in 10,000 reshufflings of the sampled ballot-polling data are shown.}
    \label{tab:kalamazoo}
\end{table}

\subsection{A highly stratified audit}

As mentioned in Section~\ref{sec:p_vals}, many within-stratum risk-measuring functions do not yield tractable expressions for $P_F(\bs{\theta})$ or $P_M(\bs{\theta})$ as a function of $\bs{\theta}$,
making it hard to find the maximum $P$-value over the union unless $K$ is small.
Indeed, previous implementations of SUITE only work for $K = 2$. 
However, the combined log-$P$-value for EB is linear in $\bs{\theta}$ for $P_M^*$ and piecewise linear for $P_F^*$.
Maximizing the combined log-$P$-value over the union of intersections is then 
a linear program that can be solved efficiently even when $K$ is large.

To demonstrate, we simulated a stratified ballot-polling audit 
of the 2020 presidential election in California, in which 
$N = 17,500,881$ ballots were cast across $K = 58$ counties (the strata),
using a risk limit of 5\%.
The simulations assumed that the reported results were correct, and checked whether reported winner Joseph R.\ Biden really beat reported loser Donald J.\ Trump. 
The audit assumed that every ballot consisted of one card; workloads would be proportionately higher if the sample were drawn from a collection of cards that includes some cards that do not contain the contest.
Sample sizes were set to be proportional to turnout, plus 10 cards, ensuring that at least 10 cards were sampled from every county.
Risk was measured within strata by EB with predictable mixture $\lambda_{ik}$ thresholded at $0.9$ \citep{WaudbysmithRamdas20}. 
Within-stratum $P$-values were combined using $P_F^*$ ($P_M^*$ did not work well for EB with proportional allocation in simulations).
To approximate the distribution of sample sizes needed to stop, we simulated 30 audits at each increment of 5,000 cards from 5,580 to 100,580 cards.
We then simulated 300 audits at 70,580 cards, roughly the 90th percentile according to the smaller simulations.

In 91\% of the 300 runs, the audit stopped by the time 70,580 cards had been drawn statewide.
Drawing 70,580 ballots by our modified proportional allocation rule produces within-county sample sizes ranging from 13 (Alpine County, with the fewest voters) to 17,067 (Los Angeles County, with the most). 
A comparison or hybrid audit using sampling without replacement would presumably require inspecting substantially 
fewer ballots.
It took about 3.5~seconds to compute each $P$-value in R (4.1.2) using a linear program solver from the \texttt{lpSolve} package (5.6.15) on a mid-range laptop (2021 Apple Macbook Pro).

\section{Discussion}
\label{sec:discussion}



ALPHA intersection supermartingales were most efficient compared to the SUITE pilot audit in Michigan and in simulations. 
Lower-sided testing allocation was better than proportional allocation, especially for EB. 
Fisher pooling limits the damage that a poor allocation rule can do, but is less efficient than intersection supermartingales with a good stratum selection rule.
For comparison audits, it helps to bet more aggressively than ALPHA-ST by using ALPHA-UB or EB. 
However, EB was not efficient compared to SUITE when replicating the Michigan hybrid audit due to poor performance in the ballot-polling stratum.

Our general recommendation for hybrid audits is: (i)~use an intersection supermartingale test with (ii)~adaptive stratum selection and (iii)~ALPHA-UB (or another method that can exploit low sample variance to bet more aggressively) as the risk-measuring function in the comparison stratum and (iv)~ALPHA-ST (or a method that ``learns'' the population mean) as the risk-measuring function in the ballot-polling stratum. 
When the number of strata is large, audits can leverage the log-linear form of the EB supermartingale to quickly find the maximum $P$-value, as illustrated by our simulated audit spread across California's 58 counties.



In future work, we hope to construct better stratum allocation rules and characterize (if not construct) optimal rules. The log-linear structure of the EB supermartingale may make it simpler to derive optimal allocation rules.

While stratum selection is not an instance of a traditional
multi-armed bandit (MAB) problem, there are connections, and successful strategies for MAB might help. 
For instance, stratum selection could be probabilistic and involve continuous exploration and exploitation, in contrast to the ``hard stop'' rules we used in our simulations here. 


\bibliography{pbsBib.bib}

\begin{thebibliography}{23}
\providecommand{\natexlab}[1]{#1}
\providecommand{\url}[1]{\texttt{#1}}
\expandafter\ifx\csname urlstyle\endcsname\relax
  \providecommand{\doi}[1]{doi: #1}\else
  \providecommand{\doi}{doi: \begingroup \urlstyle{rm}\Url}\fi

\bibitem[Appel and Stark(2020)]{appelStark20}
A.~Appel and P.~Stark.
\newblock Evidence-based elections: Create a meaningful paper trail, then
  audit.
\newblock \emph{Georgetown Law Technology Review}, 4.2:\penalty0 523--541,
  2020.
\newblock
  \url{https://georgetownlawtechreview.org/wp-content/uploads/2020/07/4.2-p523-541-Appel-Stark.pdf}.

\bibitem[Appel et~al.(2020)Appel, DeMillo, and Stark]{appelEtal20}
A.~Appel, R.~DeMillo, and P.~Stark.
\newblock Ballot-marking devices cannot assure the will of the voters.
\newblock \emph{Election Law Journal, Rules, Politics, and Policy}, 2020.
\newblock \url{https://papers.ssrn.com/sol3/papers.cfm?abstract_id=3375755}.

\bibitem[Baker and Haberman(2020)]{baker20}
P.~Baker and M.~Haberman.
\newblock In {Torrent} of {Falsehoods}, {Trump} {Claims} {Election} {Is}
  {Being} {Stolen}.
\newblock \emph{The New York Times}, Nov. 2020.
\newblock ISSN 0362-4331.
\newblock URL
  \url{https://www.nytimes.com/2020/11/05/us/politics/trump-presidency.html}.

\bibitem[Chaitlin(2020)]{chaitlin20}
D.~Chaitlin.
\newblock Sidney {Powell} shares 270-page binder of documents buttressing
  election fraud claims, Dec. 2020.
\newblock URL
  \url{https://www.washingtonexaminer.com/news/sidney-powell-shares-election-fraud-claims}.
\newblock Section: News.

\bibitem[Hall et~al.(2009)Hall, Miratrix, Stark, Briones, Ginnold, Oakley,
  Peaden, Pellerin, Stanionis, and Webber]{hallEtal09}
J.~Hall, L.~Miratrix, P.~Stark, M.~Briones, E.~Ginnold, F.~Oakley, M.~Peaden,
  G.~Pellerin, T.~Stanionis, and T.~Webber.
\newblock Implementing risk-limiting post-election audits in {C}alifornia.
\newblock In \emph{Proc.~2009 Electronic Voting Technology Workshop/Workshop on
  Trustworthy Elections (EVT/WOTE~'09)}, Montreal, Canada, August 2009.
  {USENIX}.
\newblock URL
  \url{http://www.usenix.org/event/evtwote09/tech/full_papers/hall.pdf}.

\bibitem[Higgins et~al.(2011)Higgins, Rivest, and Stark]{higginsEtal11}
M.~Higgins, R.~Rivest, and P.~Stark.
\newblock Sharper p-values for stratified post-election audits.
\newblock \emph{Statistics, Politics, and Policy}, 2\penalty0 (1), 2011.
\newblock URL \url{http://www.bepress.com/spp/vol2/iss1/7}.

\bibitem[Howard et~al.(2019)Howard, Rivest, and Stark]{howardEtal19}
L.~Howard, R.~Rivest, and P.~Stark.
\newblock A review of robust post-election audits: Various methods of
  risk-limiting audits and {B}ayesian audits.
\newblock Technical report, Brennan Center for Justice, 2019.
\newblock
  \url{https://www.brennancenter.org/sites/default/files/2019-11/2019_011_RLA_Analysis_FINAL_0.pdf}.

\bibitem[Howard et~al.(2021)Howard, Ramdas, McAuliffe, and Sekhon]{Howard21}
S.~R. Howard, A.~Ramdas, J.~McAuliffe, and J.~Sekhon.
\newblock Time-uniform, nonparametric, nonasymptotic confidence sequences.
\newblock \emph{The Annals of Statistics}, 49\penalty0 (2), apr 2021.
\newblock \doi{10.1214/20-aos1991}.
\newblock URL \url{https://doi.org/10.1214%2F20-aos1991}.

\bibitem[Kahn(2020)]{kahn20}
C.~Kahn.
\newblock Half of {Republicans} say {Biden} won because of a 'rigged' election:
  {Reuters}/{Ipsos} poll.
\newblock \emph{Reuters}, Nov. 2020.
\newblock URL
  \url{https://www.reuters.com/article/us-usa-election-poll-idUSKBN27Y1AJ}.

\bibitem[Levine(2020)]{levine20}
A.~Levine.
\newblock Donald {Trump}’s {Favorite} {Voting} {Machines}, Sept. 2020.
\newblock URL
  \url{http://washingtonmonthly.com/2020/09/23/donald-trumps-favorite-voting-machines/}.

\bibitem[Ottoboni et~al.(2018)Ottoboni, Stark, Lindeman, and
  McBurnett]{ottoboniEtal18}
K.~Ottoboni, P.~Stark, M.~Lindeman, and N.~McBurnett.
\newblock Risk-limiting audits by stratified union-intersection tests of
  elections ({SUITE}).
\newblock In \emph{Electronic Voting. E-Vote-ID 2018. Lecture Notes in Computer
  Science}. Springer, 2018.
\newblock \url{https://link.springer.com/chapter/10.1007/978-3-030-00419-4_12}.

\bibitem[Pesarin and Salmaso(2010)]{pesarinSalmaso10}
F.~Pesarin and L.~Salmaso.
\newblock \emph{Permutation tests for complex data: Theory, applications, and
  software}.
\newblock John Wiley and Sons, Ltd., West Sussex, UK, 2010.

\bibitem[Stark(2008{\natexlab{a}})]{stark08a}
P.~Stark.
\newblock Conservative statistical post-election audits.
\newblock \emph{Ann. Appl. Stat.}, 2:\penalty0 550--581, 2008{\natexlab{a}}.
\newblock URL \url{http://arxiv.org/abs/0807.4005}.

\bibitem[Stark(2008{\natexlab{b}})]{stark08d}
P.~Stark.
\newblock A sharper discrepancy measure for post-election audits.
\newblock \emph{Ann. Appl. Stat.}, 2:\penalty0 982--985, 2008{\natexlab{b}}.
\newblock URL \url{http://arxiv.org/abs/0811.1697}.

\bibitem[Stark(2009{\natexlab{a}})]{stark09a}
P.~Stark.
\newblock {CAST}: Canvass audits by sampling and testing.
\newblock \emph{IEEE Transactions on Information Forensics and Security,
  Special Issue on Electronic Voting}, 4:\penalty0 708--717,
  2009{\natexlab{a}}.

\bibitem[Stark(2009{\natexlab{b}})]{stark09c}
P.~Stark.
\newblock Auditing a collection of races simultaneously.
\newblock Technical report, {arXiv.org}, 2009{\natexlab{b}}.
\newblock URL \url{http://arxiv.org/abs/0905.1422v1}.

\bibitem[Stark(2019)]{stark19b}
P.~Stark.
\newblock Delayed stratification for timely risk-limiting audits.
\newblock \url{https://www.stat.berkeley.edu/~stark/Preprints/delayed19.pdf},
  2019.

\bibitem[Stark(2020)]{stark20a}
P.~Stark.
\newblock Sets of half-average nulls generate risk-limiting audits: {SHANGRLA}.
\newblock \emph{Financial Cryptography and Data Security, Lecture Notes in
  Computer Science}, 12063, 2020.
\newblock Preprint: \url{http://arxiv.org/abs/1911.10035}.

\bibitem[Stark(2022)]{stark22}
P.~Stark.
\newblock {ALPHA}: Audit that learns from previously hand-audited ballots.
\newblock \emph{Annals of Applied Statistics}, Conditionally accepted, 2022.
\newblock Preprint: \url{https://arxiv.org/abs/2201.02707}.

\bibitem[Stark and Wagner(2012)]{starkWagner12}
P.~Stark and D.~Wagner.
\newblock Evidence-based elections.
\newblock \emph{IEEE Security and Privacy}, 10:\penalty0 33--41, 2012.
\newblock
  \url{https://www.stat.berkeley.edu/~stark/Preprints/evidenceVote12.pdf}.

\bibitem[Ville(1939)]{ville39}
J.~Ville.
\newblock \emph{\'{E}tude critique de la notion de collectif}.
\newblock 1939.
\newblock URL \url{http://eudml.org/doc/192893}.

\bibitem[Waudby-Smith and Ramdas(2020)]{WaudbysmithRamdas20}
I.~Waudby-Smith and A.~Ramdas.
\newblock Estimating means of bounded random variables by betting, 2020.
\newblock URL \url{https://arxiv.org/abs/2010.09686}.

\bibitem[Waudby-Smith et~al.(2021)Waudby-Smith, Stark, and
  Ramdas]{waudby-smithEtal21}
I.~Waudby-Smith, P.~Stark, and A.~Ramdas.
\newblock {RiLACS}: {Risk} {Limiting} {Audits} via {Confidence} {Sequences}.
\newblock In R.~Krimmer, M.~Volkamer, D.~Duenas-Cid, O.~Kulyk, P.~Rønne,
  M.~Solvak, and M.~Germann, editors, \emph{Electronic {Voting}}, pages
  124--139, Cham, 2021. Springer International Publishing.
\newblock ISBN 978-3-030-86942-7.

\end{thebibliography}

\appendix

\section{Computational details}
\label{sec:appendix_computation}

The following describes details of the allocation simulations in Section~\ref{sec:simulations}. 
Within each stratum, we computed null means along an equispaced grid of $(2\max\{N_1,N_2\})$ points\footnote{%
The cardinality was chosen so that a null mean was computed for every possible (discrete) value of $\theta_k$. 
A finer grid is unnecessary; a coarser grid may not find the true minimum.} 
for $\theta_1 \in [\varepsilon_1, \theta/w_1 - \varepsilon_1]$ with $\theta_2 = (\theta - w_1 \theta_1)/w_2$. 
The null means were then adjusted to $\beta_1 := \theta_1 + 1 - \bar{A}_1^c$ and $\beta_2 := \theta_1 + 1 - \bar{A}_2^c$. 
The conditional null means $\beta_{i1}$ and $\beta_{i2}$ were computed as:
$$\beta_{ik} = \frac{N_k \beta_k - \sum_{j=1}^{i-1} X_{ik}}{N_k - (i - 1)}$$

Tuning parameters for ALPHA-ST were chosen as in 
\citet[Section~2.5.2]{stark22} with $d_k = 20$ and the initial estimate $\tau_{0k}$ set to $u^A_k = 1$, the expected mean when there is no error in the CVRs. 
For ALPHA-UB, we set
$$\tau_{ik}^{\tiny \mbox{UB}} := \frac{(d_k \tau_{0k} + \sum_{j=1}^{i-1} X_{jk})/(d_k + i - 1) + f_k u_k / \hat{\sigma}^2_{ik}}{1 + f_k / \hat{\sigma}^2_{ik}}.$$
The first term in the numerator of $\tau_{ik}^{\tiny \mbox{UB}}$ is truncated shrinkage estimator ALPHA-ST. 
The second term biases $\tau_{ik}^{\tiny \mbox{UB}}$ towards $u_k$ with a weight proportional to the inverse running sample variance $\hat{\sigma}^2_{ik}$. 
The constant of proportionality $f_k$ is a tuning parameter set to $f_k := .01$; higher $f_k$ would bias $\tau_{ik}$ towards $u_{k}$ more aggressively. 
The variance-dependent bias amounts to betting more when the population variance is low, which it tends to be in comparison audits when the voting system works properly. 
Truncation keeps $\tau_{ik}$ within its allowed range. 

For both ALPHA strategies, $\tau_{ik}$ was truncated to be in $[\beta_{ik} + \varepsilon_k, u_k (1 - \delta)]$, where $\varepsilon_k := {1}/{2 N_k}$ was the minimum value of one assorter and $\delta = 2.220446\times10^{-16}$ was machine precision.
If $\beta_{ik} + \varepsilon_k \geq u_{k}$, we set the corresponding terms in the supermartingale to 1: that (composite) null is true.

Each stratum selection rule was applied to every supermartingale.
For proportional allocation, there was no additional selection: samples were gathered round-robin across strata, omitting any strata that were fully exhausted. For lower-sided testing, the sampling from a stratum ceased when the lower-sided test rejected at level .05. 
This was implemented by setting all future terms in the supermartingale equal to 1 after rejection. 
The stratumwise supermartingales were then multiplied to produce $2 \max\{N_1, N_2\}$  intersection supermartingales and their minimum (over nulls) was found at each sample size. 
The reciprocal of this minimized intersection supermartingale was a sequence of $P$-values corresponding to $P_M^*$ under a particular sample allocation rule. The same strategy, but using Fisher pooling, was used to find $P_F^*$. The sample size at risk limit $\alpha = 5\%$ is the sample size for which the $P$-value sequence first hits or crosses 0.05, summed across both strata.

\section{Data and code}

All code used in this paper is available at \url{https://github.com/spertus/sweeter-than-SUITE}. 
SUITE was applied to the Michigan RLA data in a Jupyter notebook available at \url{https://github.com/kellieotto/mirla18}. 
Reported results from California's 2020 presidential election are available at 
\url{https://elections.cdn.sos.ca.gov/sov/2020-general/sov/csv-candidates.xlsx}.

\end{document}